\documentclass[conference]{IEEEtran}

\IEEEoverridecommandlockouts % This is necessary for the \thanks command to work

\usepackage{booktabs}    % For professional-quality tables
\usepackage{array}       % For better table alignment

\usepackage[table,xcdraw]{xcolor}  % For row colors
\usepackage{amsthm}
\usepackage{etoolbox}
\usepackage{textcomp}
\def\BibTeX{{\rm B\kern-.05em{\sc i\kern-.025em b}\kern-.08em
    T\kern-.1667em\lower.7ex\hbox{E}\kern-.125emX}}
\IEEEoverridecommandlockouts

\usepackage{float} % For better control of floating environments


\usepackage[normalem]{ulem}
\usepackage{cancel}
\usepackage{hyperref}
\usepackage{epsfig,epsf,epstopdf}
\graphicspath{ {./Figs/} }
\usepackage[cmex10]{amsmath}
\usepackage{amsfonts}
\usepackage{amssymb}
\usepackage[noend]{algpseudocode}
\makeatletter
\def\BState{\State\hskip-\ALG@thistlm}
\makeatother
\usepackage{cite}
\usepackage{paralist}
\usepackage{color}
\usepackage{array}
\usepackage{mathrsfs}
\usepackage{multirow}

\usepackage{graphicx}

\usepackage{nomencl}
\usepackage{framed}
\makeatletter
\renewenvironment{proof}[1][\proofname]{\par
  \normalfont \topsep6\p@\@plus6\p@\relax
  \trivlist
  \item[\hskip\labelsep
        {\bfseries\itshape
    #1:}\@addpunct{}]\ignorespaces
}{%
  \endtrivlist\@endpefalse
}
\makeatother

\newtheoremstyle{boldhead}
  {} % Space above
  {} % Space below
  {\normalfont} % Body font - non-italic
  {} % Indent amount
  {\bfseries\itshape} % Theorem head font - bold and italic
  {.} % Punctuation after theorem head
  { } % Space after theorem head
  {\thmname{#1}\thmnumber{ #2}\thmnote{ \normalfont(#3)}} 
\newtheorem{theorem}{Theorem}

\newtheorem{assumption}{Assumption}
\newtheorem{definition}{Definition}
\newtheorem{remark}{Remark}
\newtheorem{lemma}{Lemma}

\makenomenclature
\begin{document}
\title{Transient-Safe and Attack-Resilient Secondary Control in AC Microgrids Under Polynomially Unbounded FDI Attacks}

% \author{Yichao Wang, Mohamadamin Rajabinezhad and Shan Zuo,~\IEEEmembership{Member,~IEEE}% <-this % stops a space
% \thanks{Yichao Wang, Mohamadamin Rajabinezhad
%  and Shan Zuo are with the Department of Electrical and Computer Engineering, University of Connecticut, Storrs, CT, 06269 USA (e-mail: yichao.wang@uconn.edu; mohamadamin.rajabinezhad@uconn.edu; shan.zuo@uconn.edu). 
% }
% }

% \author{\IEEEauthorblockN{Yichao Wang, Mohamadamin Rajabinezhad and Shan Zuo}\\
% \IEEEauthorblockA{\textit{Department of Electrical and Computer Engineering} \\
% \textit{University of Connecticut}\\
% 371 Fairfield Way; U-4157
% Storrs, Connecticut 06269-4157 U.S.A. \\
% yichao.wang@uconn.edu, mohamadamin.rajabinezhad@uconn.edu, shan.zuo@uconn.edu}
% }

% \author{ \parbox{3 in}{\centering Huibert Kwakernaak*
%         \thanks{*Use the $\backslash$thanks command to put information here}\\
%         Faculty of Electrical Engineering, Mathematics and Computer Science\\
%         University of Twente\\
%         7500 AE Enschede, The Netherlands\\
%         {\tt\small h.kwakernaak@autsubmit.com}}
%         \hspace*{ 0.5 in}
%         \parbox{3 in}{ \centering Pradeep Misra**
%         \thanks{**The footnote marks may be inserted manually}\\
%        Department of Electrical Engineering \\
%         Wright State University\\
%         Dayton, OH 45435, USA\\
%         {\tt\small pmisra@cs.wright.edu}}
% }

\author{Mohamadamin Rajabinezhad, Nesa Shams, Yichao Wang, and Shan Zuo
% \thanks{This work was not supported by any funding.}
\thanks{Mohamadamin Rajabinezhad, Nesa Shams (Upcoming Student), Yichao Wang, and Shan Zuo are with the Department of Electrical and Computer Engineering, University of Connecticut, 371 Fairfield Way; U-4157 Storrs, Connecticut 06269-4157 U.S.A.  (Emails: mohamadamin.rajabinezhad@uconn.edu; nashams71@gmail.com; yichao.wang@uconn.edu; shan.zuo@uconn.edu.)}
}

% \thanks{This work was not supported by any funding}% <-this % stops a space
% \thanks{H. Kwakernaak is with Faculty of Electrical Engineering, Mathematics and Computer Science,
%         University of Twente, 7500 AE Enschede, The Netherlands
%         {\tt\small h.kwakernaak@autsubmit.com}}%
% \thanks{P. Misra is with the Department of Electrical Engineering, Wright State University,
%         Dayton, OH 45435, USA
%         {\tt\small pmisra@cs.wright.edu}}%
% }

\maketitle

\thispagestyle{empty} % Removes the page number in the first page

\begin{abstract}
This letter proposes a novel, fully distributed, transient-safe resilient secondary control strategies for AC microgrids, addressing unbounded false data injection (FDI) attacks on control input channels. Unlike existing methods that focus primarily on steady-state convergence, our approach guarantees transient safety, ensuring that system states remain within predefined safety bounds even during attack initiation—a critical aspect overlooked in prior research. Given the reduction of network inertia by increasing the penetration of inverted-based renewables, large overshooting and intense fluctuations are more likely to occur during transients caused by disturbances and cyber attacks. To mitigate these risks, the proposed control method enhance defense capabilities against polynomially unbounded FDI attacks, maintaining safe system trajectories for both frequency and voltage throughout the transient response. Through rigorous Lyapunov-based stability analysis, we formally certify the strategies to achieve uniformly ultimately bounded (UUB) convergence in frequency and voltage regulation, and active power sharing across multi-inverter-based AC microgrids.
Numerical simulation studies verify the effectiveness of the proposed control protocols, demonstrating improved system reliability, safety and resilience under adverse conditions. 
\end{abstract}
% \begin{IEEEkeywords}
% Transient safety, Resilience, Secondary control, Microgrids, Unbounded attacks.          
% \end{IEEEkeywords}
\section{Introduction}
Microgrids play a vital role in integrating distributed energy resources (DERs) and managing the variability of renewables like wind and solar. They can operate independently or alongside the main grid, combining DERs, energy storage, and loads to enhance control, efficiency, and reliability \cite{guerrero2010hierarchical,rajabinezhad2021chapter}. However, islanded microgrids face challenges due to the absence of traditional system inertia, making voltage and frequency stability more difficult to maintain. As microgrids evolve towards distributed operation, they depend on distributed control systems with local sensing and sparse communication networks~\cite{bidram2013secondary}. While these systems improve efficiency and responsiveness, they also create vulnerabilities to cyberattacks, such as false data injection (FDI) attacks, which can bypass detection and have severe consequences~\cite{jamali2023resilient}.
Several approaches have been proposed for enhancing resilient control in AC microgrids. For instance, \cite{jamali2023resilient} 
introduces a secondary frequency control method that utilizes a resilience index to counteract  state-dependent FDI cyber-attacks. Most studies consider disturbances, noise, faults, or attacks in AC microgrids as bounded signals. However, recent works \cite{liu2023resilient,yz052802,zhou2023distributed,wang2024secondary} show that attackers, leveraging quantum computing's rapid capabilities, can launch unbounded false data injections. These attacks threaten the stability and integrity of cybersystems, highlighting the critical need for resilient defense strategies in AC microgrids. 
In \cite{wang2024secondary}, we proposed strategies to mitigate the effects of polynomially unbounded FDI attacks on control input channels and communication link faults. 

However, 
as far as we know, the frequency and voltage
regulation in Microgrids considering the two important performance
metrics “resiliency” and “safety” simultaneously,
remains largely unexplored. This is particularly crucial for inverter-based microgrids, where the low-inertia nature leads to significant frequency fluctuations, potentially breaching system protection thresholds. Therefore, in designing control policies for frequency and voltage regulation, it is essential to ensure that frequency and voltage remains within safe limits throughout operation, preventing excessive deviations from the desired value \cite{zhang2023novel,ma2023safe,zhao2023barrier}. The classical method, model predictive control (MPC), handles dynamic constraints but is computationally demanding, often requiring reduced model orders in MG secondary control, compromising stability and safety. MPC also struggles with nonlinearities and communication issues \cite{ma2023safe}.
Recently, control barrier functions (CBFs) have emerged as an effective tool for ensuring safety in systems like collision avoidance in automated vehicles, and robotics trajectory planning \cite{ames2016control}. This inspires the development of a CBF-based safe controller for frequency and voltage regulation in microgrids \cite{zhang2023novel,ma2023safe}.
% {\color{green} Stability and safety can be classified into steady-state and transient-state \cite{7}. While transient stability in microgrid (MG) control, ensuring state trajectories converge to equilibrium, is well-studied, transient safety is often overlooked. Transient safety ensures critical states remain within safe limits during transitions, which is crucial for system reliability, especially with reduced inertia causing larger overshoots and fluctuations. Traditionally, steady-state safety is addressed at the tertiary level \cite{8}, but with increasing disturbances, transient safety must also be considered at the faster secondary control level \cite{9,10}. This paper focuses on addressing both transient stability and safety in MG secondary control.}
The letter's contributions are outlined as follows:

$\bullet$ We propose a novel, fully distributed transient-safe and attack-resilient secondary control strategies for AC microgrids that are designed to ensure both safety and resilience against polynomially unbounded FDI attacks, even during transient response. These strategies incorporate a compensational signal with adaptively tuned parameters based on neighborhood relative information, effectively addressing both frequency and voltage regulation. Unlike previous methods which primarily focus on steady-state convergence and only could handle a limited range of unbounded attacks with bounded first-order time derivatives \cite{liu2023resilient, yz052802, zhou2023distributed}, our approach guarantees that system states remain within predefined safety bounds during transient responses even under wider range of unbounded FDI attacks signals by only requiring bounded higher-order time derivatives, which is less restrictive compared with
the similar requirement found in the existing literature. This is crucial for mitigating large overshooting and intense fluctuations due to reduced network inertia and cyber attacks.

$\bullet$ We conduct a rigorous Lyapunov-based stability analysis to prove that our strategies ensure uniformly ultimately bounded (UUB) convergence for frequency, voltage regulation, and power sharing. The analysis confirms that the strategies maintain transient safety even under polynomially unbounded FDI attacks, mitigate adverse effects, and manage fault impacts effectively. By adjusting adaptation gains, we can make the ultimate bounds for frequency and voltage arbitrarily small, enhancing safety and stability.

$\bullet$ The fully distributed defense strategies require no global information, ensuring scalability and easy integration. Their effectiveness in maintaining transient safety and stability is validated through simulations on a modified IEEE 34-bus test feeder system with inverter-based resources, demonstrating significant improvements in reliability and transient-safety under adverse conditions.

\section{Preliminaries on Graph Theory and Notations}

%\subsection{Preliminaries}
Consider a network with \(N\) inverters and one leader node, which operates autonomously due to the absence of incoming edges. Followers process information from adjacent agents. The network is represented by \(\mathscr{G} = (\mathcal{V}, \mathcal{E}, \mathcal{A})\), where \(\mathcal{V}\) is the node set, \(\mathcal{E} \subset \mathcal{V} \times \mathcal{V}\) is the edge set, and \(\mathcal{A} = [a_{ij}] \in \mathbb{R}^{N \times N}\) is the adjacency matrix with \(a_{ij}\) as the edge weight: \(a_{ij} \neq 0\) if \((v_j, v_i) \in \mathcal{E}\), otherwise \(a_{ij} = 0\). There are no repeated edges or self-loops, so \(a_{ii} = 0\). Define \(\mathcal{D} = \operatorname{diag}({d_i}) \in \mathbb{R}^{N \times N}\) and \(\mathcal{L} = \mathcal{D} - \mathcal{A}\) as the in-degree and Laplacian matrices, respectively, where \(d_i = \sum_{j=1}^N a_{ij}\). Pinning gain \(g_{i}\) represents the influence of the leader on the \(i\)-th converter: \(g_{i} > 0\) if there is a link, \(g_{i} = 0\) otherwise. The matrix \(\mathcal{G} = \operatorname{diag}(g_{i})\) represents the pinning gains. Denote $\mathcal{L} + \mathcal{G} = \mathcal{L}_{\mathcal{G}}$. \(\mathscr{F}\) and \(\mathscr{L}\) are the sets \(\{1, 2, \ldots, N\}\) and \(\{N+1, N+2\}\). \(\mathbf{1}_N\) is a column vector of ones. \(\otimes\), \(\operatorname{diag}\{\cdot\}\), and \(\|\cdot\|\) denote the Kronecker product, block diagonal matrix, and Euclidean norm, respectively. 
\subsection{Safety Preliminaries}
In this section, we present the preliminaries of the CBF, a powerful tool for designing safe controllers. Consider the following affine nonlinear system:
\begin{equation}
 \dot{x}=f(x)+g(x) u
\label{affine control}
\end{equation}

where the functions 
$f$ and $g$ are locally Lipschitz continuous. Here, 
$x \in \mathbb{R}^n$ represents the state of the system, and 
$u \in \mathbb{R}^m$ is the control input.

\begin{definition}
\label{def: extended class $K$ function}
\cite{ames2016control}. A continuous function $\alpha:[-b, a) \rightarrow$ $[-\infty,+\infty)$ is said to be an extended class $K$ function for $a, b>$ 0 if it is strictly increasing and $\alpha(0)=0$.
\end{definition}

\begin{definition}
\label{def: forward invariant} \cite{xiao2019decentralized}. A set $\Omega$ is forward invariant for system (1) if its solutions starting at all $x\left(t_0\right) \in \Omega$ satisfy $x(t) \in \Omega$ for $\forall t \geq t_0$.
\end{definition}

\begin{definition}
\label{def: forward invariant2}\cite{ames2016control}.
Consider a set $\Omega = \left\{ x \in \mathbb{R}^n \mid h(x) \geq 0 \right\} $ for a continuously differentiable function $h: \mathbb{R}^n \to \mathbb{R} $. The function $h$ is termed a control barrier function if there exists an extended class \( \mathcal{K} \)  function $\alpha $ such that $\sup_{u \in \mathbb{R}^m} \left\{ L_f h(x) + L_g h(x) u + \alpha(h(x)) \right\} \geq 0,$ where $L_f h(x) = \frac{\partial h}{\partial x} f(x), \quad L_g h(x) = \frac{\partial h}{\partial x} g(x)$ are the Lie derivatives, and $ \alpha(\cdot) $ is an extended class $K$ function.
% Given a set $\Omega=\left\{x \in \mathbb{R}^n: h(x) \geq 0\right\}$ for a continuously differentiable function $h: \mathbb{R}^n \rightarrow R$, the function $h$ is called a control barrier function, if there exists an extended class $K$ function $\alpha$ such that $$
% \sup _{u \in R^{\mathrm{m}}}\left\{L_f h(x)+L_g h(x) u+\alpha(h(x))\right\} \geq 0 $$ where $$
% L_f h(x)=\frac{\partial h}{\partial x} f(x), \quad L_g h(x)=\frac{\partial h}{\partial x} g(x) $$ are Lie derivatives, $\alpha(.)$ is an extended class $K$ function.
\end{definition}

\begin{lemma}
\label{le: CBF control objective} (\cite{ames2016control}). For a CBF $h(x)$, define the set for $x \in \mathbb{R}^n$, $K_{c b f}(x)=\left\{u \in \mathbb{R}^m: L_f h(x)+L_g h(x) u+\alpha(h(x)) \geq 0\right\}$.
\end{lemma}

Then, any Lipschitz continuous controller $u \in K_{c b f}(x)$, will render the set $\Omega=\left\{x \in \mathbb{R}^n: h(x) \geq 0\right\}$ forward invariant for control system \eqref{affine control}.

\section{Standard  Consensus Control in Isolated AC Microgrids}

In islanded microgrids, feedback linearization decouples voltage and frequency dynamics. Secondary control \cite{bidram2013secondary} adjusts setpoints for decentralized primary control by providing frequency $\omega_{n_i}$ and voltage $V_{n_i}$ setpoints through neighbor data exchange. Using droop control, the frequency and voltage are defined as
${\omega }_i= {\omega }_{n_i} - {m_{P_i}}{{P}_i}$, and ${v}_{odi}= {V}_{n_i} - {n_{Q_i}}{{Q}_i}$. The dynamics of this droop-based regulation in secondary control are modeled as in \cite{bidram2013secondary, yz052802}.
\begin{align}
& {{\dot \omega }_{n_i}}={{\dot \omega }_i} + {m_{P_i}}{{\dot P}_i}=u_{f_i},
\label{eq: ufi}\\
& {{\dot V}_{n_i}}={{\dot v}_{odi}} + {n_{Q_i}}{{\dot Q}_i}=u_{v_i},
\label{eq: uvi}
\end{align}
Here, $P_i$ and $Q_i$ represent the active and reactive powers, respectively, while $\omega_i$ is the operating angular frequency, and $v_{odi}$ refers to the $d$-axis component of the inverter terminal voltage after the $abc$ to $dq0$ transformation. The droop coefficients ${m_{P_i}}$ and ${n_{Q_i}}$ correspond to the $P-\omega$ and $Q-v$ relationships, determined based on the inverters' power ratings. The auxiliary control inputs $u_{f_i}$ and $u_{v_i}$ will be designed later. The setpoints $\omega_{n_i}$ and $V_{n_i}$ are obtained by integrating $u_{f_i}$ and $u_{v_i}$ over time, which in turn regulates $\omega_i$ and $v_{odi}$ through the droop control dynamics. The local cooperative frequency and voltage control protocols at each inverter will be designed based on the following relative information with respect to the neighboring inverters and the leaders
% $\kessi$
\begin{equation}
\begin{gathered}
  \xi_{f_i} \equiv {c_f}\Big( {\sum\limits_{j \in \mathscr{F}} {a_{ij}\left( {{\omega _j} - {\omega _i}} \right)} } + {g_{i}\left( {{\omega _k} - {\omega _i}} \right)}  \hfill \\
  + \sum\limits_{j \in \mathscr{F}} {a_{ij}\left( {{m_{{P_j}}}{P_j} - {m_{{P_i}}}{P_i}} \right)}  \Big), \hfill \\ 
\end{gathered}
\label{eq: zetafi}
\end{equation}

\begin{equation}
\begin{gathered}
  \xi_{v_i} \equiv {c_v}\Big( {\sum\limits_{j \in \mathscr{F}} {a_{ij}\left( {v _{odj} - v _{odi}} \right)} } +  {g_{i}\left( {{v _k} - v _{odi}} \right)}  \hfill \\
  +\sum\limits_{j \in \mathscr{F}} {a_{ij}\left( {{n_{{Q_j}}}{Q_j} - {n_{{Q_i}}}{Q_i}} \right)}\Big), \hfill \\ 
\end{gathered}
\label{eq: zetavi}
\end{equation}
where $c_f$ and $c_v$ are constant coupling gains. Since the inverters have identical parameters, the same coupling gains are used for the neighborhood relative information at each inverter. The variables $\omega_k$ and $v_k$ represent the frequency reference and voltage reference values, respectively.
Denote ${\omega _{{n_k}}}={\omega _k} + {m_{{P_i}}}{P_i}$ and $V_{n_k}={v_k} + {n_{{Q_i}}}{Q_i}$. Utilizing the neighborhood information from Eqs. \eqref{eq: zetafi} and \eqref{eq: zetavi} \cite{bidram2013secondary}, the standard cooperative secondary control protocols for AC microgrids are expressed as ${\dot \omega }_{n_i}=u_{f_i}^c=\xi_{f_i}$ and ${\dot V}_{n_i} = u_{v_i}^c = \xi_{v_i}$.
% \begin{equation}
% \begin{gathered}
% {\dot \omega }_{n_i}=u_{f_i}=\xi_{f_i} = c_f\left(\sum\nolimits_{j \in \mathscr{F}} {{ a_{ij}}\left( {{\omega _{{n_j}}} - {\omega _{{n_i}}}} \right)}  \\
% + {{ g_{ir}}\left( {{\omega _{{n_k}}} - {\omega _{{n_i}}}} \right)}\right) ,\hfill\\
% \label{eq: frequency closed loop dynamics}
% \end{gathered}
% \end{equation}

% \begin{equation}
% \begin{gathered}
% {\dot V}_{n_i} = u_{v_i} = \xi_{v_i} =  c_v\left(\sum\nolimits_{j \in \mathscr{F}} {{a_{ij}}\left( {{V_{{n_j}}} - {V_{{n_i}}}} \right)}  \hfill\\
% + {{  g_{ir}}\left( {{V_{{n_k}}} - {V_{{n_i}}}} \right)}\right).\hfill\\
% \label{eq: voltage closed loop dynamics}
% \end{gathered}
% \end{equation}
The global vector forms of \eqref{eq: zetafi} and \eqref{eq: zetavi} are
\begin{equation}
\xi_f =  -  c_f\mathcal{L}_{\mathcal{G}}\left( {{\omega _n} - {{\mathbf{1}}_N} \otimes {\omega _{{n_k}}}} \right),
\label{eq12}
\end{equation}
\begin{equation}
\xi_v =  -  c_v \mathcal{L}_{\mathcal{G}}\left( V_n - {{\mathbf{1}}_N} \otimes {V_{n_k}} \right), 
\label{eq13}
\end{equation}
where  $\xi_f= {[ {\xi_{f_1}^T,...,\xi_{f_N}^T} ]^T}$, $\xi_v= {[ {\xi_{v_1}^T,...,\xi_{v_N}^T} ]^T}$, $\omega_n= {[ {\omega_{n_1}^T,...,\omega_{n_N}^T} ]^T}$ and $V_n= {[ {V_{n_1}^T,...,V_{n_N}^T} ]^T}$. Define the global frequency and voltage containment error vectors as
\begin{equation}
{e_f} = {\omega _n} -  (\mathcal{L}_{\mathcal{G}}^{ - 1})\mathcal{L}_{\mathcal{G}}{\left( {{{\mathbf{1}}_N} \otimes {\omega _{{n_k}}}} \right)} ,
\label{eq: ef}
\end{equation}
\begin{equation}
{e_v} = {V_n} -  (\mathcal{L}_{\mathcal{G}}^{ - 1})\mathcal{L}_{\mathcal{G}}{\left( {{{\mathbf{1}}_N} \otimes {V_{{n_k}}}} \right)}.
\label{eq: ev}
\end{equation}
\subsection{Modeling of Unbounded Attacks}
In this part, we introduce the unbounded FDI attacks on the local control inputs of the frequency and voltage control loops, then auxiliary control input signal in \eqref{eq: ufi} and \eqref{eq: uvi} becomes to:
% we examine unbounded FDI attacks on the local control inputs for the frequency and voltage control loops. We then analyze the auxiliary control input signals given in \eqref{eq: ufi} and \eqref{eq: uvi}, which, after accounting for safety constraints via the QP problem in \eqref{eq: QP2}, are transformed into:
% \vspace{-2mm}
\begin{equation}
\begin{gathered}
  \bar u_{fi} = u_{fi} + \Delta_{f_i}, \hfill 
\end{gathered}
\label{eq16}
\end{equation}
\begin{equation}
\begin{gathered}
  \bar u_{vi} = u_{vi} + \Delta_{v_i}, \hfill 
\end{gathered}
\label{eq17}
\end{equation}
where $\bar u_{fi}$ and $\bar u_{vi}$ are corrupted inputs, $\Delta_{f_i}$ and $\Delta_{v_i}$ denote the  polynomially unbounded attack signals injected to the control inputs of frequency and voltage control loops at the $i^{th}$ inverter, respectively.
\begin{assumption}
\label{ass: assumption on the attacks}
The attack signals $\Delta_{f_i}\in C^{\gamma}$ and $\Delta_{v_i}\in C^{\gamma}$ are polynomially unbounded with the finite polynomial order of $\gamma$. That is, $|{\Delta_{f_i}}^{(\gamma)}| \leqslant \kappa_{f_i}$ and $|{\Delta_{v_i}}^{(\gamma)}| \leqslant \kappa_{v_i}$, where $\gamma > 0$ is a scalar, and $\kappa_{f_i}$ and $\kappa_{v_i}$ are positive constants.
\end{assumption}

\begin{remark}
\label{rem: remark on the assumption of the attacks}
% Assumption \ref{ass: assumption on the attacks} addresses a wider range of unbounded FDI attacks, which could be deployed by adversaries exploiting the computational advantages of quantum computers. Unlike \cite{liu2023resilient,yz052802,zuo2022adaptive,zhou2023distributed}, the requirement for a bounded first-order time derivative of the attack signals is relaxed to a bounded higher-order derivative. The proposed controller is capable of managing both time-varying and constant signals, provided that $|{\Delta_{f_i}}^{(\gamma)}| \leq \kappa_{f_i}$ and $|{\Delta_{v_i}}^{(\gamma)}| \leq \kappa_{v_i}$, as specified in Assumption \ref{ass: assumption on the attacks}. It is important to note that, depending on the computational capacity of the defender, the highest derivative order, $\gamma$, can be made arbitrarily large. Moreover, polynomially unbounded attacks affecting the rate of change of controlled variables can result in rapid variations before reaching the saturation bounds, potentially causing system instability.
Assumption \ref{ass: assumption on the attacks} considers a broader range of unbounded FDI attacks, potentially exploiting quantum computational advantages. Unlike \cite{liu2023resilient,yz052802,zhou2023distributed}, it relaxes the bounded first-order derivative requirement to a bounded higher-order derivative. The proposed controller handles both time-varying and constant signals if \( |{\Delta_{f_i}}^{(\gamma)}| \leq \kappa_{f_i} \) and \( |{\Delta_{v_i}}^{(\gamma)}| \leq \kappa_{v_i} \), with \(\gamma\) being adjustable depending on the defender's computational power. Polynomially unbounded attacks may cause rapid changes, potentially leading to system instability before saturation limits are reached.
\end{remark}
\begin{definition}[\cite{yz052802}]
\label{def: UUB}
Signal $x(t)\in {\mathbb{R}^n}$ is UUB with an ultimate bound $b$, if there exist positive constants $b$ and $c$, independent of ${t_0} \geqslant 0$, and for every $a \in \left( {0,c} \right)$, there exists $t_1 = t_1 \left( {a,b} \right) \geqslant 0$, independent of $t_0$, such that $\left\| {x\left( {{t_0}} \right)} \right\| \leqslant a\Rightarrow \left\| {x\left( t \right)} \right\| \leqslant b, \forall t \geqslant {t_0} + {t_1}$.
\end{definition}

% \begin{definition}[\cite{rockafellar2015convex}]
% \label{def: convex hull}
% A set $\mathfrak{S} \subseteq \mathbb{R}^n$ is said to have its convex hull, denoted as ${\text{Co}}(\mathfrak{S})$, if it is the smallest convex set that contains all points in $\mathfrak{S}$. Formally, ${\text{Co}}(\mathfrak{S})$ is defined as:
% \[
% {\text{Co}}(\mathfrak{S}) = \left\{ \left. \sum_{i=1}^{k} \alpha_i s_i  \right| s_i \in \mathfrak{S}, \alpha_i \geq 0, \sum_{i=1}^{k} \alpha_i = 1, k \in \mathbb{N} \right\},
% \]
% where $\sum_{i=1}^{k} \alpha_i s_i$ represents the convex combination of points in $\mathfrak{S}$, ensuring that every point within ${\text{Co}}(\mathfrak{S})$ can be expressed as such a combination, thus preserving the minimal convexity criteria.
% \end{definition}

The following assumptions are needed for the communication graph topology to guarantee cooperative consensus.
\begin{assumption}
\label{ass: directed path from each leader to each inverter}
There exists a directed path from the leader to each inverter.    
\end{assumption} 
\section{Problem Formulation}
Even though the primary controller is designed to stabilize the microgrid, introducing a secondary controller can affect the system's dynamic behavior and stability. Hence, it is essential to ensure that the transient stability of the closed-loop microgrid system is maintained when the secondary controller is implemented. In this work,we are to design a control approach for frequency and voltage
regulations in microgrid while considering resilient distributed control
and safety even during the unbounded FDI attacks. 
Meanwhile, a designated safe region can be established around
the desired operating point, and we intend to ensure that the frequency and voltage of all inverters
remain within this region at all times during operation
to satisfy the goal of safety.
The transient safety bound is primarily influenced by safety considerations and the physical limitations of the hardware, and is generally larger than the steady-state bound. However, to the authors’
best knowledge, there still lacks a commonly accepted standard
suggesting the magnitude of transient safety bound for
microgrids. Consequently, we adopt the steady-state bound as the transient bound for DER output voltages and frequency as $[0.9, 1.1] \, p.u.$ and $60 \pm 2\, Hz$ respectively \cite{ma2023safe}.
Specifically, the safety constraints
for frequency and voltage are given by
\begin{equation}
-\omega_l \leq \hat\omega_i \leq \omega_h \quad , -v_l \leq \hat v_i \leq v_h \quad 
\label{eq: Constraints}
\end{equation}

where $\hat\omega_i = \omega_i - \omega_{ref}$ and $\hat v_i = v_i - v_{ref}$. Also, $\omega_l > 0$, $\omega_h > 0$, $v_l > 0$, and $v_h > 0$ which define the safety regions for frequency and voltage. 

\subsection{Safety Guarantees With Control Barrier Functions}

Throughout the operation, the frequency of each inverter must stay within the defined safety limits, which are governed by the constraints in \eqref{eq: Constraints}. Subsequently, we introduce two functions \( h_{f_{i}} \) and \( h_{v_{i}} \) as follows:
\begin{align}
h_{f_{1,i}} &= \hat\omega_i + \omega_l  \quad\quad
&&h_{f_{2,i}} = \omega_h - \hat\omega_i\\
h_{v_{1,i}} &= \hat v_i + v_l  \quad\quad
&&h_{v_{2,i}} = v_h - \hat v_i
\end{align}
Hence, to guarantee the safety of frequency is equivalent to ensuring that the following inequalities hold:
\[
h_{f_{1,i}} > 0, \quad h_{f_{2,i}} > 0, \quad h_{v_{1,i}} > 0, \quad h_{v_{2,i}} > 0,
\]

which result in a safety set:
\[
\Omega_{f_i} = \{ \hat{\omega} \mid h_{f_{1,i}} \geq 0, \; h_{f_{2,i}} \geq 0 \}
\]
\[
\Omega_{v_i} = \{  \hat{v} \mid h_{v_{1,i}} \geq 0, \; h_{v_{2,i}} \geq 0 \}
\]

As a result, when $\Omega_{f_i}$ and $\Omega_{v_i}$ are forward invariant, the frequency and voltage safety constraints are satisfied. Now, consider the control barrier functions $h_{f_{1,i}}$ and $h_{f_{2,i}}$, with their derivatives given by $h_{f_{1,i}}=\dot\omega_i$ and $h_{f_{2,i}}=-\dot\omega_i$. To guarantee the forward invariance of the set $\Omega_{f_i}$, the extended class \( \mathcal{K} \) functions \(\alpha_1\) and \(\alpha_2\) are chosen as linear functions, specifically:
\begin{equation}
\alpha_1: h_{f_{1,i}} \rightarrow \eta_1 h_{f_{1,i}}, \quad \alpha_2: h_{f_{2,i}} \rightarrow \eta_2 h_{f_{2,i}}
\end{equation}
where $\eta_1>0$ and $\eta_2>0$. For the $\Omega_{v_i}$ the procedure is the same. Following Lemma \ref{le: CBF control objective}, the conditions resulting from CBFs for safe control are given by:
\begin{equation}
 \dot{h}_{f_{1,i}} + \eta_1 h_{f_{1,i}} \geq 0, \quad \dot{h}_{f_{2,i}} + \eta_1 h_{f_{2,i}} \geq 0 \quad 
 \label{eq: CBF_Constraints}
\end{equation}

Substituting $h_{f_{1,i}}=\dot\omega_i$ and $h_{f_{2,i}}=-\dot\omega_i$ into \eqref{eq: CBF_Constraints}, and using \eqref{eq: ufi}, we see the condition \eqref{eq: CBF_Constraints} actually
contains the term $m_{pi} \dot{P}_i$, which is unknown for us. Therefore,
this condition cannot be used to search for the safe control
directly. Further, we note that the load fluctuation in the real world is not unlimited. Thus, it is reasonable to assume that
the disturbance is bounded by a positive constant $d_s$ \cite{zhang2023novel}. Then the conditions for safe control resulting from
robust control barrier functions are given by
\begin{equation}
u_{f_i} - m_{pi}d_s + \eta_1 (\omega_i + \omega_l - \omega_{ref})\geq 0,
\label{eq: safety condition1}
\end{equation}
\begin{equation}
-u_{f_i} + m_{pi}d_s + \eta_2 (\omega_h - \omega_i + \omega_{ref})\geq 0,
\label{eq: safety condition2}
\end{equation}

That is, if the control $u_{f_i}^{safe}$ is designed to satisfy conditions \eqref{eq: CBF_Constraints}, the safety can be guaranteed. The control input $u_{fi}$ is computed by solving the following optimization problem:
% \begin{equation}
% \min_{u_i \in \mathbb{R}} J(u(t)) \min_{u_i} \frac{1}{2} \| u_fi^{\text{opt}} - u_i \|^2,
% \end{equation}
\begin{equation}
% \label{eqT3.4}
\begin{aligned}
    &\quad\min_{u_{fi}^{safe} \in \mathbb{R}} J(u(t)) = \int_{t_0}^{t_f} \left( \frac{1}{2}(u_{fi}^{safe}(t) - u_{fi}^c(t))^2 \right) dt \\
    &\quad\quad\quad\quad\quad\quad\quad\text{s.t.} \quad  
    \eqref{eq: safety condition1} \quad\text{and}\quad \eqref{eq: safety condition2}
\label{eq: QP}
\end{aligned}
\end{equation}

Note that the variable in the optimization problem \eqref{eq: QP} is $u_{fi}^{safe}$, and
\eqref{eq: safety condition1} and \eqref{eq: safety condition2} are actually linear constraints. Thus the optimization
\eqref{eq: QP} is a quadratic programming (QP) problem and can be further rewritten as 

\begin{equation}
% \label{eqT3.4}
\begin{aligned}
&\arg\min_{u_{fi}^{safe}} \|u_{fi}^{safe} - u_{fi}^c\|_2 \\
&\text{s.t.} \quad  
    A_{cbf} u_{fi}^{safe}(t) \leq b_{cbf}
\label{eq: QP2}
\end{aligned}
\end{equation}

where $A_{cbf}=[-1;1]$ and $b_{cbf}=[\eta_1 (\omega_i + \omega_l - \omega_{ref}) - m_{pi}d_s ; m_{pi}d_s + \eta_2 (\omega_h - \omega_i + \omega_{ref})]$.
The optimization problem designed ensures that the control input adheres to safety constraints while minimizing deviations from the reference control input. Denote $\Delta u_{f_i}\equiv u_{fi}^{safe} - u_{fi}^{c}$. This approach enables efficient computation of the control input, upholding both system safety and resilience. The QP problem can be solved rapidly, allowing real-time control to be achieved \cite{zhang2023novel}. Due to space constraints, the voltage problem formulation is omitted, but it follows the same structure as the frequency formulation.  
\begin{remark}
\label{rem: QP bound}
In the QP problem, the objective function  $\|u_{fi} - u_{fi}^c\|$ is a measure of how close $u_{fi}$ is to $u_{fi}^c$. The boundedness of \(\|u_{fi} - u_{fi}^c\|\) is guaranteed because the QP problem is solved over a feasible set defined by the constraints \(A_{cbf} u_{fi}(t) \leq b_{cbf}\). These constraints ensure that the feasible region for \(u_{fi}\) is bounded. As a result, the minimum value of \(\|u_{fi} - u_{fi}^c\|\) is finite, making \(\Delta u_{fi}\) bounded due to the constrained, bounded nature of the optimization problem.
% The objective function  \(\|u_{fi} - u_{fi}^c\|\) is a measure of how close $u_{fi}$ is to $u_{fi}^c$. The boundedness of \(\|u_{fi} - u_{fi}^c\|\) is assured because the QP problem is defined over a feasible set constrained by \(A_{cbf} u_{fi}(t) \leq b_{cbf}\). Since these constraints define a bounded feasible region for \(u_{fi}\), the distance \(\|u_{fi} - u_{fi}^c\|\) is also bounded. Specifically, the feasible set limits the range of \(u_{fi}\), and thus, the minimum distance between \(u_{fi}\) and \(u_{fi}^c\) achievable within this bounded region will be finite. Consequently, $\Delta u_{f_i}$ is bounded because it is minimized over a constrained, bounded region. 
\end{remark} 

\subsection{Resilient control Problem Formulation}
In the face of polynomially unbounded FDI attacks, the objective of resilient frequency and voltage control is to design the control inputs $u_{f_i}$ and $u_{v_i}$ such that the local frequency $\omega_i$ and voltage $v_{odi}$ of each inverter converge to a small vicinity around the reference values set by the the leader.
\begin{lemma}[\cite{yz052802,bidram2013secondary}]
\label{lem: nonsingular and positive-definite} 
Given Assumptions \ref{ass: directed path from each leader to each inverter}, the resilient frequency and voltage consensus control objectives are achieved if $e_f \left( t \right)$ and $e_v \left( t \right)$ are UUB, respectively.
\end{lemma}

We now present the resilient defense problems for the secondary control loops governing frequency and voltage.

\begin{definition}
\label{def: Resilient Frequency and Voltage Defense Problem}
The \textbf{Resilient Frequency and Voltage Defense Problem} involves designing resilient control inputs, $u_{f_i}$ and $u_{v_i}$, as specified in Eqs. \eqref{eq: ufi} and \eqref{eq: uvi}, respectively, to ensure the resilient frequency and voltage control objectives are met. Specifically, both the global frequency containment error $e_f$, defined in Eq. \eqref{eq: ef}, and the global voltage containment error $e_v$, defined in Eq. \eqref{eq: ev}, should remain UUB despite unbounded attacks on the local frequency and voltage control input channels.
\end{definition}

% \begin{definition}
% \label{def: Resilient Frequency Defense Problem}
% The \textbf{Resilient Frequency Defense Problem} involves designing a control input $u_{f_i}^{res}$, as specified in Eq. \eqref{eq: ufi}, to ensure that the resilient frequency control objective is met. Specifically, the global frequency containment error $e_f$, defined in Eq. \eqref{eq: ef}, should remain uniformly ultimately bounded (UUB) despite unbounded attacks on the local frequency control input channels.

% \end{definition}

% \begin{definition}
% \label{def: Resilient Voltage Defense Problem}
% The \textbf{Resilient Voltage Defense Problem} focuses on designing a control input $u_{v_i}^{res}$, as detailed in Eq. \eqref{eq: uvi}, to achieve the resilient voltage control objective. Specifically, the global voltage containment error $e_v$, defined in Eq. \eqref{eq: ev}, must remain uniformly ultimately bounded (UUB) despite unbounded attacks on the local voltage control input channels.

% \end{definition}

\section{Fully Distributed Resilient Defense Strategies Design}

We propose the following fully distributed resilient defense strategies to solve the resilient frequency and voltage defense problems. we have \( u_{f_i} = u_{f_i}^{safe} + \Gamma_{f_i} \). Given $ u_{f_i}^c = \xi_{f_i} $ and  $ \Delta u_{f_i} \equiv u_{f_i}^{safe} - u_{f_i}^c$, this can be expressed as follows:

\begin{equation}
  \begin{aligned}
    &\left\{ \begin{gathered}
      u_{f_i} = {\xi _{f_i} + \Delta u_{f_i} + {\Gamma}_{f_i}}, \hfill \\
      {{\Gamma}_{f_i}} = \frac{\xi_{f_i}\Upsilon_{f_i}}{\lvert\xi_{f_i}\rvert + \eta_{f_i}}, \hfill \\ 
      {\Upsilon^{(\gamma)}_{f_i}} = \nu_{f_i}\lvert\xi_{f_i}\rvert
    \end{gathered}  \right.
    \quad
    &\left\{ \begin{gathered}
      u_{v_i} = {\xi _{v_i} + \Delta u_{v_i} +{\Gamma}_{v_i}}, \hfill \\
      {{\Gamma}_{v_i}} = \frac{\xi_{v_i}\Upsilon_{v_i}}{\lvert\xi_{v_i}\rvert + \eta_{v_i}}, \hfill \\ 
      {\Upsilon^{(\gamma)}_{v_i}} = \nu_{v_i}\lvert\xi_{v_i}\rvert
    \end{gathered}  \right.
  \end{aligned}
  \label{eq:resilient_strategies_voltage}
\end{equation}

where, $\eta_{f_i}$ and $\eta_{v_i}$ are positive, exponentially decaying functions that facilitate a smooth control strategy for practical implementation. ${\Gamma}_{f_i}$ and ${\Gamma}_{v_i}$ represent compensatory signals, while $\Upsilon_{f_i}$ and $\Upsilon_{v_i}$ are adaptively tuned parameters. The adaptation gains $\nu_{f_i}$ and $\nu_{v_i}$ are fixed positive constants, and the initial values of both $\Upsilon_{f_i}$ and $\Upsilon_{v_i}$ are also positive.

\begin{theorem}
\label{thm: resilient frequency defense strategies}
Based on Assumptions \ref{ass: directed path from each leader to each inverter} and \ref{ass: assumption on the attacks}, and applying the cooperative resilient frequency defense strategies outlined in Eq. \eqref{eq: zetafi} and Eq. \eqref{eq:resilient_strategies_voltage}, the global frequency containment error $e_f$, as defined in Eq. \eqref{eq: ef}, remains UUB, thus solving the resilient frequency defense problem. Moreover, by appropriately increasing the adaptation gain ${\nu _{f_i}}$, as indicated in Eq. \eqref{eq:resilient_strategies_voltage}, the ultimate bound of $e_f$ can be reduced to an arbitrarily small value.
\end{theorem}
\begin{proof}
Combining \eqref{eq12}, \eqref{eq16}, \eqref{eq: zetafi} and \eqref{eq:resilient_strategies_voltage} yields the vector form: 
\begin{equation}
\begin{gathered}
\dot{\xi}_f = -c_f\mathcal{L}_{\mathcal{G}}\dot \omega_n\hfill = - c_f\mathcal{L}_{\mathcal{G}}
\times\big(\xi_f +\Delta_f + \Delta u_{f} +\Gamma_f\big),\hfill
\end{gathered}
\label{eq: zeta i dot}
\end{equation}
where $\xi_f= [ \xi_{f_i}^T,...,\xi_{f_N}^T ]^T, \Delta_f= [ \Delta_{f_i}^T,...,\Delta_{f_N}^T ]^T,  \Delta u_{f}= [ \Delta u_{f_i}^T,...,\Delta u_{f_N}^T ]^T$ and $\Gamma_f= [ \Gamma_{f_i}^T,...,\Gamma_{f_N}^T ]^T$. Consider the following Lyapunov function \begin{align}
E & =\xi_f^{T}\mathcal{L}_{\mathcal{G}}^{-1} \xi_f.
\label{eq: the Lyapunov function}
\end{align}
The time derivative of \eqref{eq: the Lyapunov function} is
\begin{equation}
\begin{gathered}
\dot{E}=2 \xi_f^{T}\mathcal{L}_{\mathcal{G}}^{-1} \dot{\xi}_f\\
=\xi_f^{T}\times2\mathcal{L}_{\mathcal{G}}^{-1}\Big( - c_f\mathcal{L}_{\mathcal{G}}
\big(\xi_f +\Delta_f + \Delta u_{f} + \Gamma_f\big)\Big)\\
=  -2c_f\xi^T_f\xi_f - 2c_f\xi^T_f\Delta_f - 2c_f\xi^T \Delta u_{f} -2c_f\xi^T_f\Gamma \hfill\\
\leqslant -2c_f\sum\nolimits_{i \in \mathscr{F}}\xi^2_{f_i} + 2c_f\sum\nolimits_{i \in \mathscr{F}} \big|\xi_{f_i} \big|(\big|\Delta_{f_i}\big|+ \big|\Delta u_{f_i}\big|)\hfill\\
-2c_f\sum\nolimits_{i \in \mathscr{F}} \big(\xi_{f_i} \Gamma_{f_i}\big).\hfill
\end{gathered}
\label{eq: time derivative of the Lyapunov function}
\end{equation}

After substituting $\Gamma_{f_i}$ as defined in  \eqref{eq:resilient_strategies_voltage}, Eq. \eqref{eq: time derivative of the Lyapunov function} becomes
\begin{align}
& \dot{E} \leqslant -2c_f\sum\limits_{i \in \mathscr{F}}\xi^2_{f_i} + 2c_f\sum\limits_{i \in \mathscr{F}}\bigg(\big|\xi_{f_i} \big|(\big|\Delta_{f_i}\big|+ \big|\Delta u_{f_i}\big|)  \nonumber\\
&-\Big(\big|\xi_{f_i}\big|^2 \times\Upsilon_{f_i}\Big)/\big(\left|\xi_{f_i}\right|+\eta_{f_i}\big) \bigg)
\nonumber\\
& = -2c_f\sum\limits_{i \in \mathscr{F}}\xi^2_{f_i} + 2c_f\sum\limits_{i \in \mathscr{F}}\bigg(\big|\xi_{f_i}\big|^2(\big|\Delta_{f_i}\big| + |\Delta u_{f_i}\big|)-\big|\xi_{f_i}\big|^2\nonumber\\
&\times\Upsilon_{f_i}   + \big|\xi_{f_i} \big|\big|\Delta_{f_i}\big|\eta_{f_i} + \big|\xi_{f_i} \big|\big|\Delta u_{f_i}\big|\eta_{f_i}  \bigg)/\big(\left|\xi_{f_i}\right|+\eta_{f_i}\big).
\label{eq: the second part of the derivation of the time derivative of the Lyapunov function}
\end{align}

Per Assumption~\ref{ass: assumption on the attacks} and note that $\eta_{f_i}$ is a exponentially decaying function and the decay rate of $\eta_{f_i}$ exceeds the growth rate of polynomially unbounded signals, we have $\lim_{t \to \infty}  \big(\big|\xi_{f_i} \big|\big|\Delta_{f_i}\big|\eta_{f_i}\big) = \lim_{t \to \infty}  \big(\big|\xi_{f_i} \big|\big|\Delta u_{f_i}\big|\eta_{f_i}\big)= 0$. Furthermore, noting that ($-2c_f\sum\nolimits_{i \in \mathscr{F}}\xi^2_{f_i}$) is negative, and $\Delta u_{f_i}$ is bounded and can be disregarded compared to $\Delta_{f_i}$, which is polynomially unbounded signal, design $\Upsilon_{f_i}$ as in \eqref{eq:resilient_strategies_voltage}, then when $ \left|\xi_{f_i}\right| > \kappa_{f_i} / \nu_{f_i}$, $\exists t_1$, such that
\begin{align}
& 2c_f\sum\limits_{i \in \mathscr{F}}\bigg(\big|\xi_{f_i}\big|^2(\big|\Delta_{f_i}\big| + |\Delta u_{f_i}\big|)-\big|\xi_{f_i}\big|^2\nonumber\times\Upsilon_{f_i}    \bigg)\\
&  /\big(\left|\xi_{f_i}\right|+\eta_{f_i}\big)\leqslant 0,\,\forall t \geqslant t_1.
\label{eq: critical parts of time derivative of the Lyapunov function}
\end{align}

Define the compact set $\Psi_i\equiv\ \{|\xi_{f_i}|\leqslant\kappa_{f_i}/ \nu_{f_i}\}$. Considering \eqref{eq: the second part of the derivation of the time derivative of the Lyapunov function} and \eqref{eq: critical parts of time derivative of the Lyapunov function}, we obtain that outside the compact set $\Psi_i$, $\dot E \leqslant 0,\,\forall t\geqslant t_1$. By LaSalle’s invariance principle \cite{krstic1995nonlinear}, we obtain that $\xi_f$ is UUB. From Theorem 4.18 of \cite{yz053105}, the upper bound is a class $\mathcal{K}$ function of $\kappa_{f_i}/ \nu_{f_i}$. Therefore, the larger the value of the adaptation gain $\nu_{f_i}$, the smaller the ultimate bound. Note that $\xi_f = -c_f \mathcal{L}_{\mathcal{G}} e_f$. Hence $e_f$ is also UUB. This completes the proof.
\end{proof}
\begin{remark}
\label{rem: remark on the high control gain issue.}
Although, the larger the values of $\nu_{f_i}$, the smaller the ultimate bound, excessively high gains $\nu_{f_i}$ lead to overshoot, instability, control accuracy reduction, etc. Moderate gains balance responsiveness and stability, ensuring effective control.    
\end{remark}

\begin{theorem}
\label{thm: resilient voltage defense strategies}
Given Assumptions \ref{ass: directed path from each leader to each inverter} to \ref{ass: assumption on the attacks}, and using the cooperative resilient voltage defense strategies delineated by Eq. \eqref{eq: zetavi} and Eq. \eqref{eq:resilient_strategies_voltage}, the global voltage containment error $e_v$ defined in Eq. \eqref{eq: ev}, is UUB, i.e., the resilient voltage defense problem is solved. Furthermore, by properly incrementing the adaptation gain \(\nu_{v_i}\) in \eqref{eq:resilient_strategies_voltage}, the ultimate bound of \(e_v\) can be rendered arbitrarily small.
\end{theorem}

\begin{proof}
Note that $\xi_v = -c_v \mathcal{L}_{\mathcal{G}} e_v$. The approach used to prove Theorem \ref{thm: resilient voltage defense strategies} mirrors the one of Theorem \ref{thm: resilient frequency defense strategies}.
\end{proof} 

\section{Validation and Results}
\begin{figure}[ht]
\centering
\includegraphics[width=3in]{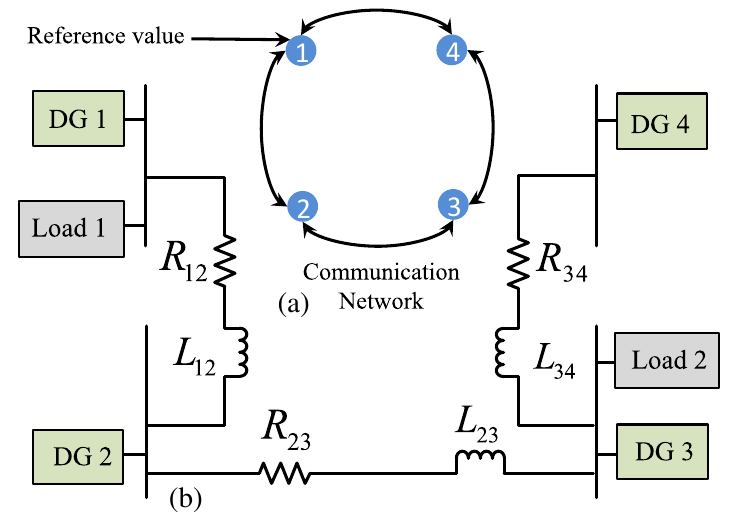}
\caption{Cyber-physical microgrid system: (a) Communication graph topology among four inverters and two leaders (references), (b) Microgrid test setup.}
\label{FIG3}
\end{figure}

% \begin{figure}[!t]
% \centering
% \includegraphics[width=3.20in]{safety_Amin.drawio.pdf}
% \caption{Control system topology}
% \label{FIG3090}
% \end{figure}
A 50-Hz islanded, three-phase inverter-based AC microgrid is utilized to assess the proposed fully distributed, transient-safe resilient secondary defense strategies. Fig.~\ref{FIG3} shows the microgrid setup, which consists of four inverter-based distributed generators (DGs) and two loads. The detailed dynamics of each inverter, described in \cite{bidram2013secondary} and \cite{zuo2016distributed}. The parameters for a microgrid test system are as follows: \( m_{P} \) and \( n_{Q} \) are \( 9.4 \times 10^{-5} \) and \( 1.3 \times 10^{-3} \) for DGs 1 and 2, and \( 18.8 \times 10^{-5} \) and \( 2.6 \times 10^{-3} \) for DGs 3 and 4. Transmission line parameters are: resistances \( R_{12} = 0.23 \, \Omega \), \( R_{23} = 0.35 \, \Omega \), \( R_{34} = 0.23 \, \Omega \), and inductances \( L_{12} = 318 \, \mu H \), \( L_{23} = 847 \, \mu H \), \( L_{34} = 318 \, \mu H \). Loads have resistances of \( 3 \, \Omega \) and inductances of \( 6.4 \, mH \) for Load 1 and \( 12.8 \, mH \) for Load 2. The proposed control defense strategy is demonstrated in Matlab Simulink through two case studies. The inverters communicate on a bidirectional communication network with the adjacency matrix of $\mathcal{A}=[0~1~0~1;1~0~1~0;0~1~0~1;1~0~1~0]$. The pinning gains are $g_{15}=1$. The frequency and voltage reference, are $60 \,\operatorname{Hz}$, and  $340\,\operatorname{V}$, respectively.
\setlength{\belowcaptionskip}{-10pt}
% \setlength{\belowcaptionskip}{default}
% \begin{table}[H]
% \centering
% \caption{Parameters of the Microgrid Test System}
% \begin{tabular}{@{}>{\bfseries}l>{\centering}p{3cm}>{\centering\arraybackslash}p{3cm}@{}}
% \toprule
% \textbf{Parameter} & \textbf{DGs 1 and 2} & \textbf{DGs 3 and 4} \\ \midrule
% \( m_{P} \) & \( 9.4 \times 10^{-5} \) & \( 18.8 \times 10^{-5} \) \\
% \( n_{Q} \) & \( 1.3 \times 10^{-3} \) & \( 2.6 \times 10^{-3} \) \\ \midrule
% \end{tabular}

% \begin{tabular}{@{}>{\bfseries}l>{\centering}p{2cm}>{\centering}p{2cm}>{\centering\arraybackslash}p{2cm}@{}}
% \toprule
% \textbf{Component} & \textbf{Lines 1-2} & \textbf{Lines 2-3} & \textbf{Lines 3-4} \\ \midrule
% \( R \) (\(\Omega\)) & \( R_{12} = 0.23 \) & \( R_{23} = 0.35 \) & \( R_{34} = 0.23 \) \\
% \( L \) (\(\mu H\)) & \( L_{12} = 318 \) & \( L_{23} = 847 \) & \( L_{34} = 318 \) \\ \midrule
% \end{tabular}

% \begin{tabular}{@{}>{\bfseries}l>{\centering}p{3cm}>{\centering\arraybackslash}p{3cm}@{}}
% \toprule
% \textbf{Loads} & \textbf{Load 1} & \textbf{Load 2} \\ \midrule
% \( R_{L} \) (\(\Omega\)) & \( 3 \) & \( 3 \) \\
% \( L_{L} \) (\(mH\)) & \( 6.4 \) & \( 12.8 \) \\ \bottomrule
% \end{tabular}
% \label{tab:microgrid_parameters}
% \end{table}

\subsection{Case Study I: Resilience Against the Unbounded Attacks without safety constraints}

\begin{figure}[t]
\centering
{\includegraphics[width=3in]{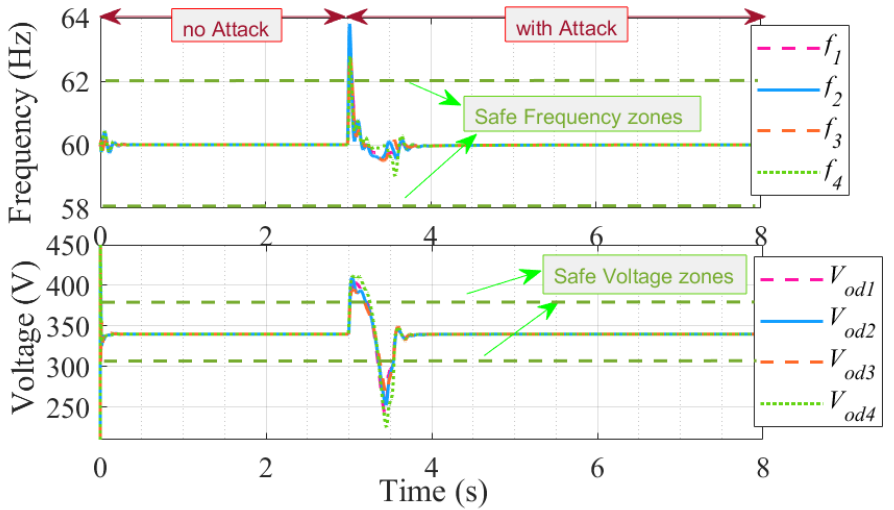}}
\caption{Performance of the resilient fully distributed resilient secondary defense strategies in Case I:  frequency and voltage performance.}
\label{FIG50}
\end{figure}

\begin{figure}[t]
\centering
{\includegraphics[width=3in]{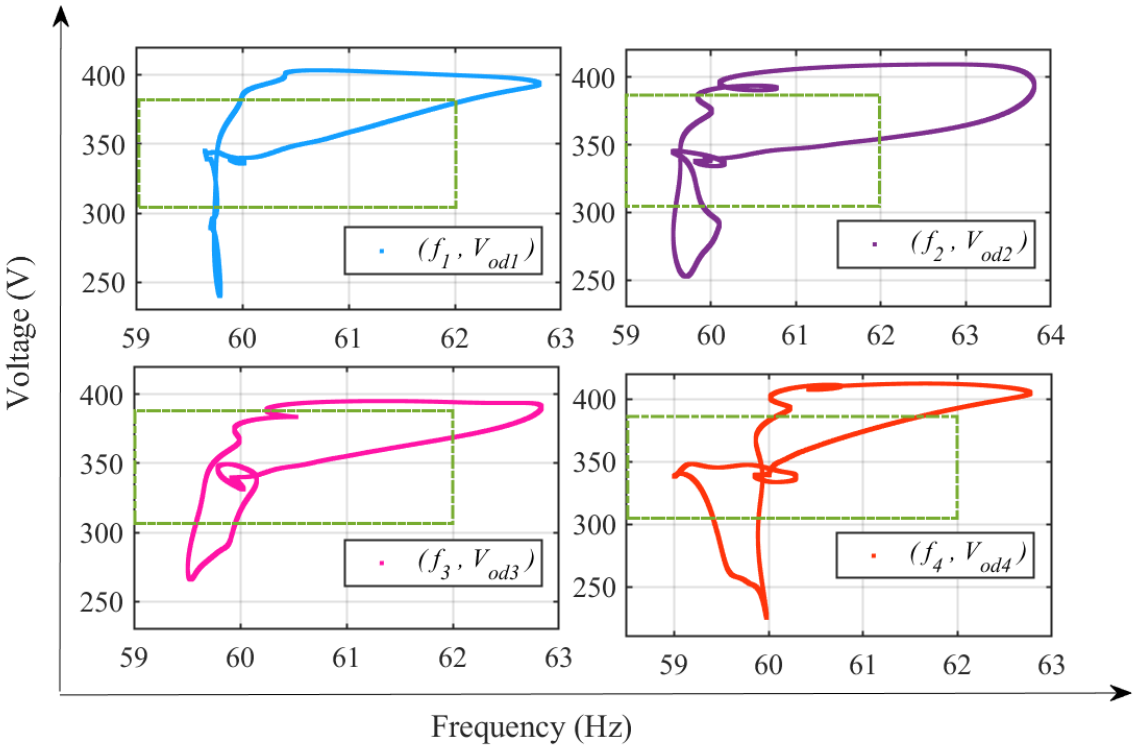}}
\caption{Trajectory of DG's output voltages and frequency usning the resilient control strategy without transient-safety.}
\label{FIG500}
\end{figure}

In this case study, the unbounded attack injections to the frequency and voltage control loops are $\Delta_{f_1}=2t^2+10, \Delta_{f_2}=2.5t^2+12, \Delta_{f_3}=-1.5t^2+6, \Delta_{f_4}=-3t^2-12$ and $\Delta_{v_1}=1.5t^2+50, \Delta_{v_2}=3t^2+15, \Delta_{v_3}=-2t^2+30, \Delta_{v_4}=2t^2+50$, respectively.
The performance of the resilient defense strategies, as defined in \eqref{eq:resilient_strategies_voltage}, is evaluated in the condition with out considering the transient-safety constraints. The constant gains $c_f=20, c_v=10$, and the adaptation gains for the resilient defense strategies are set at ${\nu_{v_i}}=20,  {\nu_{f_i}}=350, i=1,2,3,4.$ $\eta_{v_i}$ and $\eta_{f_i}$ are determined as $e^{-\alpha_{v_i}}$ and $e^{-\alpha_{f_i}}$, respectively, where ${\alpha_{v_i}}={ \alpha_{f_i}}=0.01, i=1,2,3,4$. Figure \ref{FIG50} contrast the voltage and frequency of each DGs responses to the unbounded attacks using resilient defense strategies without considering the safety constraints. Also Figure \ref{FIG500} indicate the phase plot of the trajectory of voltage and frequency during the simulation time span. The results indicate that while the resilient secondary defense strategies yield favorable outcomes—such as inverter voltages stabilizing near \(340\,\mathrm{V}\) and frequency aligning with the \(60\,\mathrm{Hz}\) reference—the transient response of both voltage and frequency violates the safety limits after unbounded FDI attacks are initiated at \(t = 5\,\mathrm{s}\), due to the absence of safety constraints.
% \begin{figure}[!ht]
% \centering
% {\includegraphics[width=3.3in]{Fig2.png}}
% \caption{Performance of the conventional secondary control in Case I: frequency, active power, and voltage performance.}
% \label{FIG40}
% \end{figure}
\setlength{\belowcaptionskip}{-10pt}
\subsection{Case Study II: Transient-Safe and Resilience Against the Unbounded Attacks}
In the second case study the safety constraints also being considered. The proposed transient-safety resilient secondary defense strategies lead to favorable results: inverter voltages stabilize within small neighbour of $340\,\mathrm{V}$ and frequency aligns with the $60\,\mathrm{Hz}$ reference. Besides, the frequency and voltage at all buses always keep
in the safe regions during the whole process including the initiation of unbounded FDI attacks on control input. These outcomes validate the proposed resilient and safe control approach's ability to ensure UUB convergence in frequency and voltage regulation in multi-inverter-based AC microgrids under polynomially unbounded attacks and keep the voltage and frequency in a predifiende safe region all during the process operation.
\setlength{\belowcaptionskip}{-10pt}
\begin{figure}[t]
\centering
{\includegraphics[width=3in]{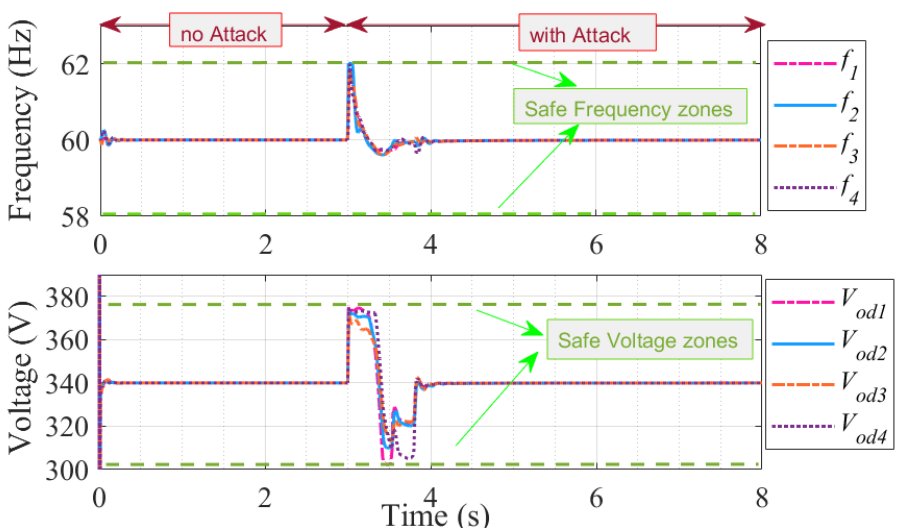}}
\caption{Performance of the proposed fully distributed transient-safety resilient  resilient secondary defense strategy in Case II: frequency and voltage performance.}
\label{FIG100}
\end{figure}
\setlength{\belowcaptionskip}{-10pt}
\begin{figure}[ht]
\centering
{\includegraphics[width=3in]{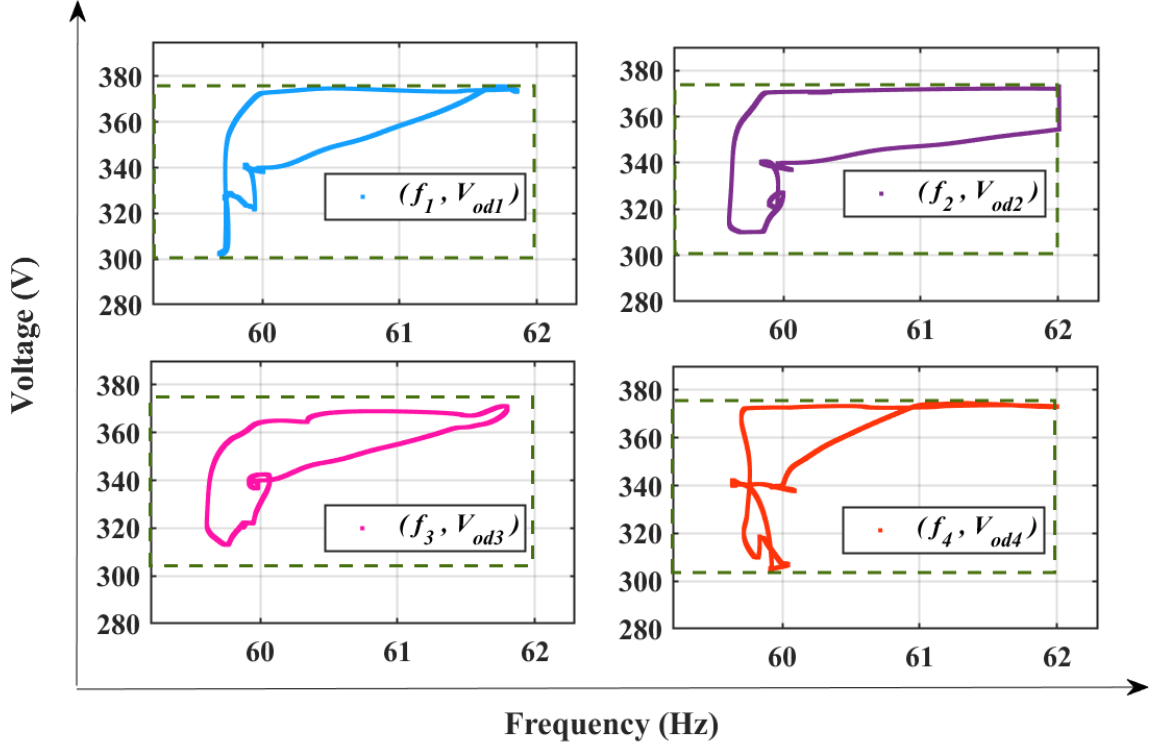}}
\caption{Trajectory of DG's output voltages and frequency usning the proposed transient-safe and resilient control strategy}
\label{FIG1000}
\end{figure}
\section{Conclusion}
This letter presents novel, fully distributed, transient-safe and attack-resilient secondary control strategies for AC microgrids that address polynomially unbounded FDI attacks on control input channels. Unlike existing methods that only guarantee steady-state resilience for limited range of unbounded attacks, our approach ensures transient safety by keeping system states within predefined bounds during wider range of unbounded attacks and disturbances. It enhances defense against risks associated with reduced network inertia caused by increasing penetration of inverted-based renewables, such as overshooting and fluctuations. The strategies ensure UUB convergence for frequency and voltage regulation, and active power sharing, with stability proven through Lyapunov analysis. Adaptation gains, $\nu_{f_i}$ and $\nu_{v_i}$, can be tuned to refine UUB stability. Simulations on a modified IEEE 34-bus system demonstrate improved reliability, safety, and resilience under adverse conditions.
 
% \ifCLASSOPTIONcaptionsoff
%   \newpage
% \fi
\bibliographystyle{IEEEtran}

\bibliography{References}

\end{document}